\documentclass[showpacs,preprint,aps]{revtex4}
\usepackage{graphicx}
\usepackage{dcolumn}
\usepackage{bm}
\begin{document}
\setcounter{page}{1}
\title
{Breit type equation for mesonic atoms}
\author
{N. G. Kelkar and M. Nowakowski}
\affiliation{ Departamento de Fisica, Universidad de los Andes,
Cra.1E No.18A-10, Santafe de Bogota, Colombia}
\begin{abstract}
The finite size effects and relativistic corrections 
in pionic and kaonic hydrogen are evaluated by
generalizing the Breit equation for a spin-0 - spin-1/2 amplitude with 
the inclusion of the hadron electromagnetic form factors.
The agreement of the relativistic corrections to the energies of 
the mesonic atoms with other methods used to evaluate them is not exact, but
reasonably good.
The precision values of the energy shifts due to the strong interaction,  
extracted from data, are however subject to the 
hadronic form factor uncertainties.
\end{abstract}
\maketitle
More than fifty years after their first appearance \cite{camac}, hadronic
atoms continue to be important for a better understanding of fundamental
interactions. One of the first speculations of their existence came from the
historic papers of Fermi, Teller and Wheeler\cite{wheel}, where they showed 
that the time required for a negative meson (they were actually referring to
muons) to be trapped into an atomic
orbit would be ($\sim 10^{-13}$ s) much less than its mean weak decay
lifetime ($\sim 10^{-6}$ s). The negative hadron which is generally trapped
into an excited state, undergoes transitions to lower states until it 
eventually enters the field of the nuclear strong interaction. The energy
levels and widths of the hadronic atomic states are naturally affected by 
the strong interaction and hence experimental programmes to measure the 
shifts in the energies and widths in pionic \cite{piexp,piothers}, kaonic 
\cite{dear}, $\Sigma$-hyperonic, antiprotonic and pionium 
\cite{sigexp} atoms accurately are being carried out 
vigorously with
the aim of pinning down the strong interaction parameters. However, the 
extraction of these parameters to a good accuracy, requires the determination of
the electromagnetic corrections accurately too. For example, the availability
of precision data on pionic hydrogen \cite{piexp} and 
deuterium \cite{pidatom} has led to 
calculations of various electromagnetic corrections to the hadronic 
scattering lengths to better than 1$\%$ \cite{sigg,eric}.

Whilst most of the 
calculations in recent literature aim at a high accuracy in evaluating 
corrections such as those due to vacuum polarization, relativistic recoil and
other higher order corrections, the finite size of the pion and the proton 
is treated in a rather simplistic way. The correction to the binding energy 
of the pionic hydrogen, due to the extended charge of the pion and the proton 
is given in some works as \cite{lyubov}, 
\begin{equation}\label{simplecorr}
\Delta E = {2 \over 3}\, \mu^3\,\,\alpha^4\,\,\biggl[ \langle r_{\pi}^2 
\rangle \,+\, \langle r_p^2 \rangle \biggr] \, ,
\end{equation}
where, $\mu$ is the reduced mass of the $\pi p$ system, $r_{\pi}$ and 
$r_p$ the charge radii of the $\pi$ and $p$ respectively and $\alpha$ the
usual fine structure constant. In \cite{sigg}, the Coulomb potential
was modified by introducing a Gaussian charge distribution which 
depended on the pion and proton charge radii. In the present work, we 
evaluate the relativistic and 
finite size corrections (FSC) by modifying the Breit equation 
\cite{lali4} to include the meson (pion or kaon) and the proton electromagnetic 
form factors. 
For similar Breit-like approaches, see \cite{breitlike}.
The results of this Breit type 
equation approach are compared with an 
`improved Coulomb potential' \cite{austen} which has been used in 
\cite{sigg,piexp}, to
obtain corrections due to relativistic recoil and the anomalous magnetic 
moment of the proton, while extracting the strong energy shift in pionic 
hydrogen. 
Using the available parameterizations of the hadron form factors, 
we also investigate the uncertainty in the estimate of the FSC. 
Considering 
the high precision with which the strong energy shifts and widths  
for the $1s$ pionic hydrogen states, namely, 
$\epsilon_{1s} = -7.108 \pm 0.013$ (stat) $\pm 0.034$ (syst) eV and 
$\Gamma_{1s} = 0.868 \pm 0.040$ (stat) $\pm 0.038$ (syst) eV \cite{piexp}, as 
well as the hadronic $\pi N$ scattering length, $a_{\pi^- p}^h = 0.0870(5)$ 
m$_{\pi}^{-1}$ \cite{eric} are being quoted and the accuracy with which the 
one loop calculations for the ground state energy of the pionic hydrogen 
are carried out \cite{gasser}, the present results become relevant. 

There is no unique approach to calculate relativistic 
corrections to level shifts of bound two-body systems 
\cite{breitlike,austen,pilkuhn1}.
We shall employ the technique of the Breit equation as it
is particularly suited to include form-factors effects in a rather
transparent way. This way an equation emerges which combines
relativistic and finite size (FSC) effects. A further motivation
to use the Breit approach is to compare it with results obtained
in a different way. Regarding the relativistic corrections, it is known
that the Breit equation is consistent at the order $\alpha^4$
\cite{pilkuhn2} and using first order time-independent perturbation theory
to calculate the energy corrections \cite{pilkuhn3}. The presence of negative
energy states poses a problem in perturbation theory at higher orders
\cite{pilkuhn3}.
A detailed comparison between the results obtained in the Breit
framework and an equation which correctly projects the positive energies
has been performed in \cite{salp}. The correction to the Breit energy 
in this work is given as, $\Delta E_{cc}=-2\alpha^5\mu^3/3\pi M_{\pi} M_p$ 
($\mu$ is the reduced mass, $M_{\pi}$ the pion and $M_p$ the 
nucleon mass) which applied to pionic atoms gives $6 \times 10^{-5}$ eV. 
This is too small to be of relevance here.  
 
To evaluate the complete electromagnetic potential, we expand the 
amplitude for $\pi p$ elastic scattering, in $1/c^2$ terms, thereby
generalizing the Breit type equation \cite{lali4} by the inclusion
of the proton and pion electromagnetic form factors. 
This leads to non-local terms in the potential, whose 
contributions are not negligible \cite{weold}.  
The $p\gamma p$ and the $\pi^-  \gamma  \pi^-$ vertices 
can be written in terms of the form factors 
$F_1^p$, $F_2^p$ (representing the charge and magnetization distributions 
in the proton) and 
$F^{\pi}$ (charge distribution in the pion) as,
\begin{eqnarray}\label{vertices}
\Gamma_p^{\mu} &= &F_1^p\, \gamma^{\mu}\,\, - \, \, {\sigma^{\mu \nu} 
\over 2 M_p\, c}\,q_{\nu}\, F_2^p \\ \nonumber
\Gamma_{\pi}^{\nu} &= &F^{\pi}(q^2) \,(P_2 + P_2^{\prime})^{\nu}\, ,
\end{eqnarray}
The photon four-momentum, $q = P_1^{\prime} - P_1 = P_2 - P_2^{\prime}$. In 
the non-relativistic limit ($q^0 = 0$) and $q^2 = - \vec{Q}^2$, where, 
$\vec{Q} = \vec{p}_1^{\,\,\prime} - \vec{p}_1 = \vec{p}_2 - 
\vec{p}_2^{\,\, \prime}$.
\begin{figure}[h]
\includegraphics[width=8cm,height=4cm]{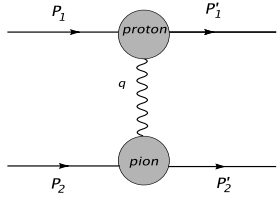}
\caption{\small Feynman diagram for pion-proton scattering}
\end{figure}
The amplitude for the process $ \pi^- + p \rightarrow \pi^- + p$ 
is given by \cite{weold},
\begin{equation}\label{ampli1}
M_{fi} = - \alpha [ \, ( \bar{u}(\vec{p}_1^{\,\,\prime})\, 
\Gamma_p^{\mu}\, u(\vec{p}_1) )\, \, D_{\mu \nu}(\vec{Q})\,\, 
\Gamma_{\pi}^{\nu}]\,
(1 / \sqrt{2E_2/c} \,)\, (1 / \sqrt{2E_2^{\prime}/c}\,),
\end{equation}
where, 
$D_{\mu \nu}(\vec{Q})$ is the photon propagator and 
$u(\vec{p}_1)$, $u(\vec{p}_1^{\,\prime})$, the Dirac spinors given as,
$u =\sqrt{2M}[
(1 - {\vec{p}^{\,2} \over 8 M_p^2c^2} ) \,w), 
{\vec{\sigma} \cdot \vec{p} \over 2 M_p c} \, w)]^T$. 
Substituting for the $u(\vec{p})$'s and the vertex factors, 
$\Gamma^{\mu}_p$ and $\Gamma^{\nu}_{\pi}$, the 
amplitude in (\ref{ampli1}) 
is evaluated and then rearranged to be written in the form, 
\begin{equation}\label{ampli2}
M_{fi} = - 2 \,M_p\,[w_1^{\prime *}\,\,
U(\vec{p}_1, \vec{p}_2, \vec{Q})\,w_1\,]\,,
\end{equation}
thus obtaining the potential in momentum 
space: 
$U(\vec{p}_1,\vec{p}_2,\vec{Q}) = - 4 \pi\,\alpha\,\sum_i \,\,
U_i(\vec{p}_1,\vec{p}_2,\vec{Q})$, where, 
\begin{eqnarray}\label{potmom}
U_1 & =& {F_1^p(Q^2)\,F^{\pi}(Q^2) \over Q^2}\,, \,\,\, 
U_2 = - \,{F^{\pi}(Q^2) \, F_2^p(Q^2)\over 4 \,M_p^2\,c^2}\,,\,\,
U_3 = - \,{F_1^p(Q^2)\,F^{\pi}(Q^2) \over M_p \,M_{\pi}\,c^2\,Q^2}\,
\biggl [ \, \vec{p}_1 \cdot \vec{p}_2 \biggr ]\,, \nonumber \\
U_4 &=&  \,{F_1^p(Q^2)\,F^{\pi}(Q^2) \over M_p \,M_{\pi}\,c^2\,Q^2}\,
\,\biggl [ {(\vec{p}_1 \cdot \vec{Q})\, 
(\vec{p}_2 \cdot \vec{Q}) \over Q^2}\, \biggr ]\,,\,\,
U_5 =  - \,{F^{\pi}(Q^2) \, F_1^p(Q^2)\over 8 \,M_p^2\,c^2}\\ \nonumber
U_6 &=& - \, {F_1^p(Q^2)\,F^{\pi}(Q^2) 
\over 2 \,M_p\,M_{\pi}\,c^2\,Q^2}\,[\,i\,(
\vec{\sigma }_1 \times \vec{Q}) \cdot \vec{p}_2\,] \,, \,\,\,
U_7 = - {F_2^p(Q^2)\,F^{\pi}(Q^2) \over 
2\,M_p \, M_{\pi}\, c^2\,Q^2 }\,[i\,(\vec{\sigma }_1 \times \vec{Q})
\cdot\vec{p}_2]\\ \nonumber
U_8 &=& {F_1^p(Q^2)\,F^{\pi}(Q^2) \over 4\, M_p^2\, c^2\,Q^2}\,
[i\,\vec{\sigma}_1\,\cdot (\vec{Q} \times \vec{p}_1)] \, , \,\,\,
U_9 = {F_2^p(Q^2)\,F^{\pi}(Q^2) \over 2\,M_p^2\,c^2\, 
Q^2}\,[i\,(\vec{p}_1 \times \vec{\sigma}_1) \cdot \vec{Q}]\,.
\end{eqnarray}
The potential in $r$-space is got by Fourier transforming 
each of the above terms \cite{lali4}, namely,
\begin{equation}\label{potr}
V_i(\vec{p}_1, \vec{p}_2,\vec{r}) \,=\, \int \,e^{i\vec{Q}\cdot\vec{r}}\,
\,U_i(\vec{p}_1, \vec{p}_2, \vec{Q}) 
\,\,{d^3Q \over (2\,\pi)^3}\,\,.
\end{equation}
The vectors $\vec{p}_1$ and $\vec{p}_2$ become 
differential operators in $r$-space \cite{lali4}.  
The FSC to the $1s$ state in pionic hydrogen can now be calculated as,
$\Delta E = {\alpha \over a} \, + \, \sum_i \Delta E_i $, where, 
\begin{equation}\label{energ}
\Delta E_i = \int \Psi_{100}(r) \,\, V_i(\vec{p}_1, \vec{p}_2, r)\,\, 
\Psi_{100}(r) \,d\vec{r}
\end{equation}
and $\Psi_{100}(r) = e^{-r/a}/(\sqrt{\pi \, a^3})$ with the Bohr radius 
$a = 1/\alpha \mu$ and $\mu$ the $\pi p$ reduced mass. 
The factor 
$\alpha /a$ in $\Delta E$, 
arises from the fact that the potential $V_1(\vec{p}_1, \vec{p}_2, r)$ 
contains the 
usual ($1/r$) Coulomb potential too which must be subtracted while 
calculating $\Delta E_1$. The spin-dependent terms ($U_6$ to $U_9$) do not
contribute to $\Delta E$ for the $1s$ state (see the appendix of \cite{weold}). Expressing $F_1^p$ and $F_2^p$,  
in terms of the Sachs form factors \cite{weold}, 
$G_E^p(Q^2)$ and $G_M^p(Q^2)$ and using Eqs (\ref{potmom}, \ref{potr} and 
\ref{energ}), the total $\Delta E$ is given as a sum of the terms: 
\begin{eqnarray}\label{allcorrs}
\Delta E_{12} &=& \biggl (\,{\alpha \over a} \,+\, 
\Delta E_1 \,\biggr )\,+\, \Delta E_2 = -{32 \,\alpha \over \pi\,a^4}\, 
\int_0^{\infty}\,{F^{\pi}(Q^2)\,G_E^p(Q^2)\,dQ \over (a^{\prime 2}+Q^2)^2}
\,+\,{\alpha \over a}
\\ \nonumber
\Delta E_{34} &=& \Delta E_3 + \Delta E_4 = {16\,\alpha \over \pi\,
M_p\,M_{\pi}\,a^5}\,\int_0^{\infty}\,{F_1^p(Q^2)\,F^{\pi}(Q^2) \over Q^2}\, 
\biggl [\,{tan^{-1}(Qa/2) \over Q}\, - \,{a^{\prime}  \over 
a^{\prime 2} + Q^2}\,\biggr ] \,dQ \\ \nonumber
\Delta E_5 &=& {4\,\alpha \over \pi\, M_p^2\,a^4}\,\int_0^{\infty}\, 
F_1^p(Q^2)\,F^{\pi}(Q^2)\, {Q^2 \over (a^{\prime 2} + Q^2)^2}\,dQ\,,
\end{eqnarray}
with, $a^{\prime} = 2/a$. 
In the above, the individual contribution due to $\Delta E_2$ is found to
be much smaller than $(\alpha /a) + \Delta E_1$ 
and in fact the two expressions can be
combined to be written in the above compact form for $\Delta E_{12}$. The
term $U_3$ in (\ref{potmom}) gives rise to
the correction $\Delta E_3$ which cancels exactly with part of the term 
arising from $\Delta E_4$ and hence we write the total sum of these two terms
above as $\Delta E_{34}$. Putting $F_1^p = G_E^p = F^{\pi} = 1$ in 
(\ref{allcorrs}), i.e. in the case of point hadrons, one gets 
from (\ref{allcorrs}) the Coulomb term plus relativistic corrections.

We use two forms for the form factors of the proton. 
In the standard dipole form,
$G_E^p(Q^2) \simeq G_M^p(Q^2) / \mu_p \simeq G_D(Q^2)$,
with, $G_D (Q^2) = 1 / (1 \,+\, Q^2/m^2)^2 $,  
$m^2 = 0.71$ GeV$^2$ and $\mu_p$ the magnetic moment
of the proton. The other parameterization is one of the 
latest phenomenological fit \cite{walch}, where, 
$G_E^p$ and $G_M^p$ are given by the ansatz, 
$G_N(Q^2) = G_S(Q^2) + a_b \, Q^2\,G_b(Q^2)$. 
The explicit forms of $G_S(Q^2)$ and $G_b(Q^2)$ are given in \cite{walch} and 
the parameters for the proton form factors are given in Table II of
\cite{walch}. 
The existing data \cite{amendolia,piffdata} 
on the pion form factor is well reproduced by a 
monopole form, namely, 
\begin{equation}
F^{\pi}(Q^2) \,= \,{1 \over 1 \,+ \, (\langle r_{\pi}^2 \rangle / 6)\,Q^2}\, 
=\, {\Lambda_{\pi}^2 \over \Lambda_{\pi}^2 \,+\,Q^2}
\end{equation}
such that, $\Lambda_{\pi}^2 = 6/\langle r_{\pi}^2 \rangle$. 
The corrections (\ref{allcorrs}) 
can be evaluated analytically, 
using the dipole form of the proton form factors. 
Since the analytic expressions for $\Delta E$ are lengthy, we give below, 
only the leading terms (in $\alpha$) 
of each of these terms. The sum of the corrections using $U_1$ and $U_2$ 
is denoted as $\Delta E^{lead}_{12}$, that 
coming from $U_3$ and $U_4$ 
in (\ref{potmom}) as $\Delta E^{lead}_{34}$ and one arising due to 
$U_5$ as $\Delta E^{lead}_5$. 
\begin{eqnarray}\label{leading}
\Delta E^{lead}_{12} &=&\biggl ({16 \,\alpha\,m^4\,\Lambda_{\pi}^2\, \over a^3}\, 
\biggr )\,\,\biggl [{2\,M_p^2\,d_1 \over m_{13}
} \,+\,{3 M_p^2\,d_2 \over m_{13}^2} 
 \,-\,{\Lambda_{\pi}^2 \, 
d_2 \over 4 m_{13}^2} \biggr ]\, ,\\ \nonumber
\Delta E^{lead}_{34} &=& {8\,\alpha\,\Lambda_{\pi}^2\,m^4\,M_p \over 
M_{\pi} \,a^3}\,\,d_3\,-\,{4\,\alpha\,M_p\,m^4\,\Lambda_{\pi}^2 \over M_{\pi}\, 
a^3}\,\biggl[{e_1 \over \Lambda_{\pi}^2} + {e_2 \over 4 M_p^2} +
{e_3 \over m^2} + {e_4 \over m^4} \biggr ]\,,
\\ \nonumber
\Delta E^{lead}_5 &=& {2\,\alpha\,\Lambda_{\pi}^4\,m^4 \over a^3}\,{d_2 \over
m_{13}^2}\,, \nonumber \\
d_1 &=& {-1 \over m_{42}^2 m_{12}^2} + {1 \over m_{42}^2 m_{14}^2} - 
{2 \over m_{12}^3 m_{42}}\,,\,\,\,
d_2 = {-1 \over m_{42}^2 m_{12}} + {1 \over m_{42}^2 m_{14}}  - {1 \over
 m_{42} m_{14}^2} \nonumber \\
d_3 &=& {1 \over m_{23} m_{43}^2 m_{31}} + {1 \over m_{23} m_{42}^2 m_{12}} + {
1 \over m_{42}^2 m_{43} m_{41}} + { 1 \over m_{42} m_{43}^2 m_{41}} + { 1 
\over m_{42} m_{43} m_{41}^2}\nonumber \\
e_1 &=& {1 \over m_{43}^2 m_{23}}\,, \,\,e_2 = {-1 \over m_{42}^2 m_{23}}\,,
\,\,e_3 = {m_{42} + m_{43} \over (m_{42} m_{43} )^2}\,, 
\,\,e_4 = {1 \over m_{42} m_{43}}
\end{eqnarray}
where we denote, $m_{ij} = m_i - m_j$, with $m_1 = a^{\prime 2}$, 
$m_2 = 4 M_p^2$, $m_3 = \Lambda_{\pi}^2$ and $m_4 = m^2$. 
Recall that $a = 1/\alpha \mu$, $a^{\prime} = 2/a$ and
hence each of the above $\Delta E$ terms are proportional to $\alpha^4$. 
Note that if one further expands the coefficients $d_1$ and $d_2$ to retain
only the leading terms, one indeed recovers Eq. (\ref{simplecorr}) 
from $\Delta E^{lead}_{12}$ above. 
The calculations using the recent parameterization of \cite{walch} are 
performed numerically. In Table I, we list the corrections
to the binding energy of pionic hydrogen, $\Delta E$, using the 
two parameterizations of the proton form factors as well as two different 
values of $\Lambda_{\pi}$ in the pion form factor. The value of 
$\langle r_{\pi}^2 \rangle = 0.439$ fm$^2$ is obtained from older 
$\pi e$ scattering experiments \cite{amendolia} and 
$\langle r_{\pi}^2 \rangle = 0.5476$ fm$^2$ is taken from a recent measurement
at the Mainz Microton facility \cite{mami}.
Although it is usually agreed that the true pion charged radius
$\langle r^2_{\pi} \rangle=0.439 \,\, {\rm fm}^2$, 
we have displayed the sensitivity
of the energy correction to the pion radius by invoking the result of a second
independent measurement \cite{mami}. As noted in \cite{mami}, 
the disagreement between
the two measurements is supposedly due to a model dependence 
in the extraction of the value of the radius.  
The error on the value of $\Delta E_{12}$ is evaluated using 
standard error propagation methods (see below). 
The contributions 
$\Delta E_{34}$ and $\Delta E_5$ are not sensitive to these errors. 

\begin{table}
\caption{\small Corrections $\Delta E$ in eV to pionic hydrogen 
using $\langle r_{\pi}^2 \rangle = 0.439$ fm$^2$. 
Numbers in brackets correspond to $\langle r_{\pi}^2 \rangle = 0.5476$ fm$^2$. 
The errors bars are due to the errors on the
proton form factors.}
\vspace{0.5cm}
\label{tab:2}
\begin{tabular}{|c|c|c|}
\hline
 &$F_1^p$, $F_2^p$ of Ref.\cite{walch}  & Dipole form \\ \hline
$\Delta E_{12}$ (eV) & 0.102${\pm 0.009}$ (0.111 ${\pm 0.009}$) 
& 0.095 (0.104)\\
$\Delta E_{34}$ (eV) & 0.0388 (0.0388)  & 0.0388  (0.0388)\\
$\Delta E_5$ (eV) & 0.0029 (0.0029)  & 0.0029 (0.0029)  \\
Total $\Delta E$ (eV) &0.143${\pm 0.009}$ (0.153${\pm 0.009}$)
 & 0.137 (0.146) \\ 
\hline
\end{tabular}
\end{table}

In what follows, we shall compare the relativistic 
corrections of the present approach with 
approaches in literature which have been used for the extraction of 
the strong interaction shift, $\epsilon_{1s}$, in pionic hydrogen, 
defined as, 
$\epsilon_{1s} = E^{e.m.}_{3p \to 1s} \,-\, E^{measured}_{3p \to 1s}$.
$E^{measured}_{3p \to 1s}$ is the measured transition energy 
\cite{piexp} and 
$E^{e.m.}_{3p \to 1s}$ is the 
calculated electromagnetic transition energy (here the strong 
interaction shift of the $3p$ state is assumed to be negligible). 
$E^{e.m.}_{3p \to 1s}$ consists of the 
energy due to the Coulomb potential between point particles and 
various electromagnetic corrections \cite{sigg}. 
Let us first consider the relativistic correction to the standard 
non-relativistic Schr\"odinger equation. 
The Hamiltonian of 
the Breit equation (in the centre of mass system, where 
$\vec{p}_1 = \vec{p}_2 = -i \vec{\nabla} = \vec{p}$) is given as, 
$H_{Breit} = \vec{p}^{\,2}/2 \mu\,-
\,\vec{p}^{\,4}/  8\, \mu^3 \,c^2\, +\, V(\vec{p},\vec{r})$.
Evaluation of the second term in the above equation, treating it as usual
\cite{grif} as a perturbation, leads to the relativistic correction to the
Bohr energy ($E^{1s}_B = - \mu \alpha^2/2$) of the $1s$ state, namely,
$\Delta E^{1s}_{rel} = - (5/8) \,\mu\,\alpha^4\,=\,- \,0.215 \,{\rm eV}$.
In the `Improved Coulomb Potential' (ICP) approach of Ref. \cite{austen}, 
starting from Eqs (9) and (10) 
in \cite{austen}, one can find the total energy, $E^{1s}_B + E^{ICP}$, for the 
case of a spin-1/2 and spin-0 bound state, where, 
\begin{eqnarray}\label{swart}
E^{ICP} 
&=& -{5 \over 8} \, \mu\, \alpha^4 \,+ \, 2 \kappa_p \biggl ( {\mu \over 
M_p} \biggr )^2 \biggl ({\mu \alpha^4 \over 2} \biggr )\, -\, 
{\mu \alpha^4 \over 2} \biggl [ {\mu \over 4 (M_p + M_{\pi})} - {2 \mu \over 
M_p + M_{\pi}} - \biggl( {\mu \over M_p}\biggr )^2\,\biggr]\, \nonumber \\
 &=& -\,0.215 \,{\rm eV}\,+\,0.01 \,{\rm eV}\,+\, 0.037\,{\rm eV}\,\,
 = \Delta E^{1s}_{rel} \,+ \,0.047 \,{\rm eV}\,.
\end{eqnarray}
Here, $\kappa_p = \mu_p - 1$, with $\mu_p = 2.793$ nm.
The first term represents the relativistic correction, which is referred to
as the standard Klein-Gordon result in \cite{austen}. This term is identical
to $\Delta E^{1s}_{rel}$ obtained from the Breit type equation. 
In the third approach, 
one could actually use the Klein-Gordon (KG) equation as was done 
in \cite{piexp,sigg}. Here the difference between the KG result, $E^{1s}_{KG}$ 
and the Bohr energy, $E^{1s}_B$ is, $E^{1s}_{KG} - E^{1s}_B 
= \Delta E^{1s}_{KG} = - 0.211$ eV.

In order to compare the terms apart from $\Delta E_{rel}$ in the Breit type 
equation
approach with those in \cite{austen}, we assume point-like hadrons such that 
the energies in (\ref{allcorrs}) become, 
\begin{equation}
\Delta \tilde{E}_{34}^{1s} + \Delta \tilde{E}_5^{1s} = 
{\alpha^4\,M_p \,M_{\pi}^3 
\over 2 \,(M_p + M_{\pi})^3}\,\biggl [\,1 \,+\, 2 {M_p \over M_{\pi}} 
\, \biggr ]  \, = \, 0.0417 {\rm eV}\,.
\end{equation}
with the tilde indicating the fact that the energies correspond to point-like 
hadrons. From Table I, we can see that $\Delta E_{34} + \Delta E_5$ is 
not different from $\Delta \tilde{E}_{34} + \Delta \tilde{E}_5$ (up to the 
fourth digit after the decimal) and the effect of the hadron form factors
on these two corrections is negligible. Besides this, we also note that 
in contrast to \cite{austen}, the 
contribution of the proton magnetic moment in the present work is found to be
negligible. This can be seen by examining Eqs (\ref{leading}) which are 
obtained analytically assuming dipole proton form factors. Though, $F_1^p$ 
in (\ref{allcorrs}) contains both the electric and magnetic Sachs 
form factors, there appears no term with $\kappa_p$ in the corrections at 
leading order in $\alpha$ as in (\ref{leading}). To summarize the above,
we have three different approaches of summing the relativistic corrections:
\begin{eqnarray}
E_{Breit}^{3p \to 1s}
 &=& \Delta E^{3p \to 1s}_{rel} \,- \,0.0417\, {\rm eV} \,=
\,0.171\,{\rm eV} \nonumber \\
E_{ICP}^{3p \to 1s}
 &=& \Delta E^{3p \to 1s}_{rel}\,-\,0.047\,=\,0.166\,{\rm eV} 
\nonumber \\
E_{Sigg}^{3p \to 1s}
 & = & \Delta E^{3p \to 1s}_{KG} \,-\,0.047\,=\,0.161\,{\rm eV}\,.
\end{eqnarray}
As can be seen there is a slight dependence on the approach used to calculate
the relativistic corrections.
It is somewhat inconsistent to use $E^{Sigg}$ \cite{piexp,sigg} 
as the sum of relativistic
corrections, since $E^{Sigg}$ is a sum of $\Delta E^{3p \to 1s}_{KG}$
and $0.047$ eV, where $0.047$ eV is taken from $E^{ICP}$
(where $\Delta E^{3p \to 1s}_{rel} \neq \Delta E^{3p \to 1s}_{KG}$). 
Using a correction of
$\Delta E_{rel}^{3p} = 0.0239$ to the Bohr energy of the $3p$ state,
namely, $E_B^{3p} = 359.441$, in Table II we present a consistent deduction
of the strong energy shift. 
The relativistic and FSC are taken from the
Breit type equation approach and the remaining corrections are as in 
\cite{piexp}. With the potential (\ref{potr}) being short-ranged, the 
finite size and relativistic corrections, $\Delta E_{12} + \Delta E_{34} 
+ \Delta E_5$, to the energy of the $3p$ state are very small and hence 
neglected.   

\begin{table}
\caption{\small Contributions in (eV) to $E^{e.m.}_{3p \to 1s}$, 
and the deduced strong interaction shift, 
$\epsilon_{1s}$ using $\langle r_{\pi}^2 \rangle = 0.439$ fm$^2$. 
Numbers in brackets correspond to ($\langle r_{\pi}^2 \rangle = 0.5476$ fm$^2$).
}
\vspace{0.5cm}
\label{tab:3}
\begin{tabular}{|c|c|}
\hline
Point Coulomb, $E_B^{3p \to 1s}$ + $\Delta E^{3p \to 1s}_{rel}$ & 2875.7196 \\ 
Breit type equation (with finite size) & -0.143$\pm 0.009$ 
(-0.153$\pm 0.009$)\\ 
Vacuum Polarization, order $\alpha^2$ \cite{piexp}
 & 3.235$\pm 0.001$ \\ 
~~~  Vacuum Polarization, order $\alpha^3$ \cite{piexp}
& 0.018 \\ 
~~~  Vertex correction  \cite{piexp} & -0.007$\pm 0.003$  \\ 
Pionic atom recoil energy \cite{piexp} &-0.004 \\ 
Total calculated $E^{e.m.}_{3p \to 1s}$ & 2878.8186 $\pm 0.009$ (2878.8086 $\pm 0.009$) \\ 
$E^{measured}_{3p \to 1s}$ \cite{piexp} & 2885.916 
$\pm 0.013 ({\rm stat})\pm 0.033 ({\rm syst})$ \\ 
$\epsilon_{1s} = E^{e.m.}_{3p \to 1s} - E^{measured}_{3p \to 1s}$ & 
-7.097 $\pm 0.009 ({\rm FSC})$ $\pm 0.013 ({\rm stat})\pm 0.033 ({\rm syst})$ 
\\
&
(-7.107 $\pm 0.009 ({\rm FSC}) \pm 0.013 ({\rm stat})\pm 0.033 ({\rm syst})$)  
\\ \hline
\end{tabular}
\end{table}

As evident from Table II, the error due to the electromagnetic form-factors 
of the proton
is of the same order as the statistical and systematic counterparts. 
Therefore some remarks on its
determination are in order. The $6 \times 6$ correlation matrix
$\rho_{ij}={\rm Cov}(a_i, a_j)/\sigma_i \sigma_j$ ($a_i$ are the fitted 
parameters and $\sigma_i$ their respective errors) 
was supplied to us by the authors of \cite{walch}. The error on $\Delta E$ 
due to uncertainties of hadronic form-factors is
calculated by the standard method, i.e.
\begin{equation} \label{error}
\left(\delta E\right)^2_{\rm FSC}= (\Delta \chi)^2
\sum_{i,j}\frac{\partial \Delta E}{\partial a_i}\biggl|_0 \,
{\rm Cov}(a_i, a_j)\frac{\partial \Delta E}{\partial a_j}\biggr |_0
\end{equation}
where the subscript $0$ denotes the central value. 
Taking $(\Delta \chi)^2=1$, we obtain the 1-$\sigma$ error on 
$\Delta E$, namely, $(\delta E)_{\rm FSC}^{1\sigma}=\pm 0.009$ eV. 
For 2-$\sigma$ variations in the parameters, $\Delta \chi^2$ increases
by 4 and the error on $\Delta E$ is doubled.

Within the framework of the present work, the correction to the 
energy of the $1s$ state in kaonic hydrogen 
(using the proton form factors of \cite{walch} and a monopole kaon form 
factor with $\langle r_K^2 \rangle = 0.34$ fm$^2$) is, 
$E^{Breit}_{kaon} = \Delta E^{1s}_{rel} + \Delta E$ (FSC) $= 0.573$ eV  
+ $2.525$ eV = $3.098 $ eV (using central values of form factor parameters). 
This correction would be relevant when better data on kaonic hydrogen would
become available from the ongoing programme of the DEAR collaboration 
\cite{dear}. 

In summary, we can say that the present work investigates 
the relativistic and finite size 
corrections in hadronic atoms, using a Breit-type equation.
These corrections 
have been shown in the present work to be important 
for the precision measurements of the strong energy shifts in pionic 
hydrogen. 
We find that the contribution of the magnetic moment
of the proton to the corrections is negligible. 
In future, we plan to extend such calculations for the evaluation of a 
spin-0 - spin-1 amplitude 
which would be relevant for the pionic deuterium case. 
The full electromagnetic potential in the $\pi d$ case will involve the 
deuteron electric, magnetic and quadrupole form factors.
In the $\pi d$ atom, the strong energy shift has been found to be 
repulsive, namely, $\epsilon_{1s} = 2.43 \pm 0.1$ eV \cite{pidatom},  
with the contribution
of the FSC, $0.51$ eV (using the simple Eq. (\ref{simplecorr}) with the 
proton radius replaced by the deuteron radius).  
The above approach could alter the precision values 
for pionic deuterium obtained so far. 
\vskip0.5cm
{\large \bf Acknowledgment}\\
The authors wish to thank Profs Th. Walcher and J. Friedrich for 
useful discussions related to evaluation of uncertainties due to errors
on the proton form factor parameters.

\end{document}